Rubbery layer on polystyrene surface well below the glassy temperature

Zongyi Qin, K. C. Wong, and Z. Yang[*]
Department of Physics, The Hong Kong University of Science and Technology, Clearwater Bay, Kowloon, Hong Kong

**Abstract**

The dynamics of polymer surfaces is a controversial subject. Many experimental evidences support the existence of a surface mobile layer, while there are a few equally strong experimental evidences that contradict it. Here we show, through careful studies of the topological evolution of nanometer-size ridges and ditches on rubbed polystyrene surfaces, that there is indeed a surface rubbery layer whose mobility decreases rapidly with depth. The segments in the layer can only be displaced by a finite distance comparable to the depth. The temperature and depth dependence of the surface effective compliance is quantitatively determined. At tens of degrees below the bulk glass transition temperature the thickness of the surface rubbery layer is only a fraction of nanometer while the rest of the polymer sample remains frozen. Such surface mobile layer can qualitatively explain the seemingly contradictory behaviors of polymer surfaces and resolve most of the controversy among the results obtained from different experimental approaches.

---

[*] Corresponding author. E-mail address: phyang@ust.hk

The dynamics of polymer surfaces is an intriguing subject in polymer physics [1 – 20], in addition of being of significant technological importance. There have been many experimental evidences that demonstrate the existence of a surface mobile layer [3 - 14]. Positron annihilation [3] and muon spin probes [4] indicated the presence of a surface mobile layer about 2 nm in thickness near the polystyrene (PS) surface. The glass transition temperature ($T_g$) of PS thin films were found to decreases drastically with the reduction of thickness [5, 6], and later studies revealed the broadening of the glass transition in polymer thin films with depth [7 – 9]. These phenomena are attributed to the existence of a liquid-like surface layer that extends deeper with the rise of temperature and diverges at bulk $T_g$. Surface friction force microscopy also showed a significant decrease in $T_g$ of the surface layers [10 – 12]. The higher mobility of the surface layer is attributed to the chain ends enrichment at the surface, which was confirmed by experiments [11]. Gold spheres of 10 nm and 20 nm in diameter placed on PS surfaces were found to sink into PS at temperatures well below bulk $T_g$ [13, 14], and the phenomenon was interpreted as due to the presence of a liquid-like surface layer.

Despite the rich evidences listed above, there are a few equally strong experimental evidences that seem to contradict the presence of such mobile layer [15 - 17]. Using a vigorous viscoelastic mechanic contact analysis, Hutcheson and McKenna showed that the sinking of gold spheres on PS surface could actually take place without the decrease in surface $T_g$ [15]. The creep compliance of 30 nm thick PS films was found to have a lower $T_g$ but much stiffened rubbery plateau than the bulk value [16]. Finally, friction force microscopy on poly(*tert*-butyl acrylate) showed no decrease in surface $T_g$ [17].

Gentle rubbing with a velvet cloth on PS surfaces can create molecular segment alignment and distortion [18 – 21], in addition to nanometer-size surface ridges and ditches. In this paper, we report the detailed study of the topological evolution of these ridges/ditches. A surface rubbery layer with enhanced mobility and depth-dependent $T_g$ is identified. Based on a phenomenological model the temperature and depth dependence of the surface effective compliance (SEC) is quantitatively determined through the analysis of experimental data. At a temperature tens of degrees below the bulk $T_g$, a surface layer only a fraction of nanometer thick is rubbery while the rest of the film remains frozen. The surface relaxation model that emerges from the analysis can qualitatively explain the seemingly contradictory behaviors of polymer surfaces [2] and resolve most of the controversy among the results obtained from different experimental approaches [3 – 15].

Figure 1 shows the topography of a PS surface just after rubbing (Fig. 1(a)), and after consecutive 30-minute annealing at 40 °C, 50°C, 60, 70°C, 80°C, 90°C, and 100°C (Fig. 1(b)). Figure 1(c) shows the line scan across the images from the same location. The deeper ditches that had survived the annealing can be matched well with the original ones before annealing. The general evolution pattern is a nearly uniform reduction of the height of the ridges and the depth of the ditches. There is no irregular local strain driven relaxation [22], indicating that the Laplace pressure (LP) is the main driving force [23]. The changes in height or depth across the entire cross section range between 1 – 2 nm, and the shallow ditches with original depth < 1 nm were smeared out. No lateral movement of the ditch/ridge walls was observed. The center positions of the ridges and ditches remained unchanged. This is consistent with the statistical aspect ratio data in Ref. 23. Figure 1(d) shows the average roughness (root mean square) of the area under investigation as a function of annealing temperature.

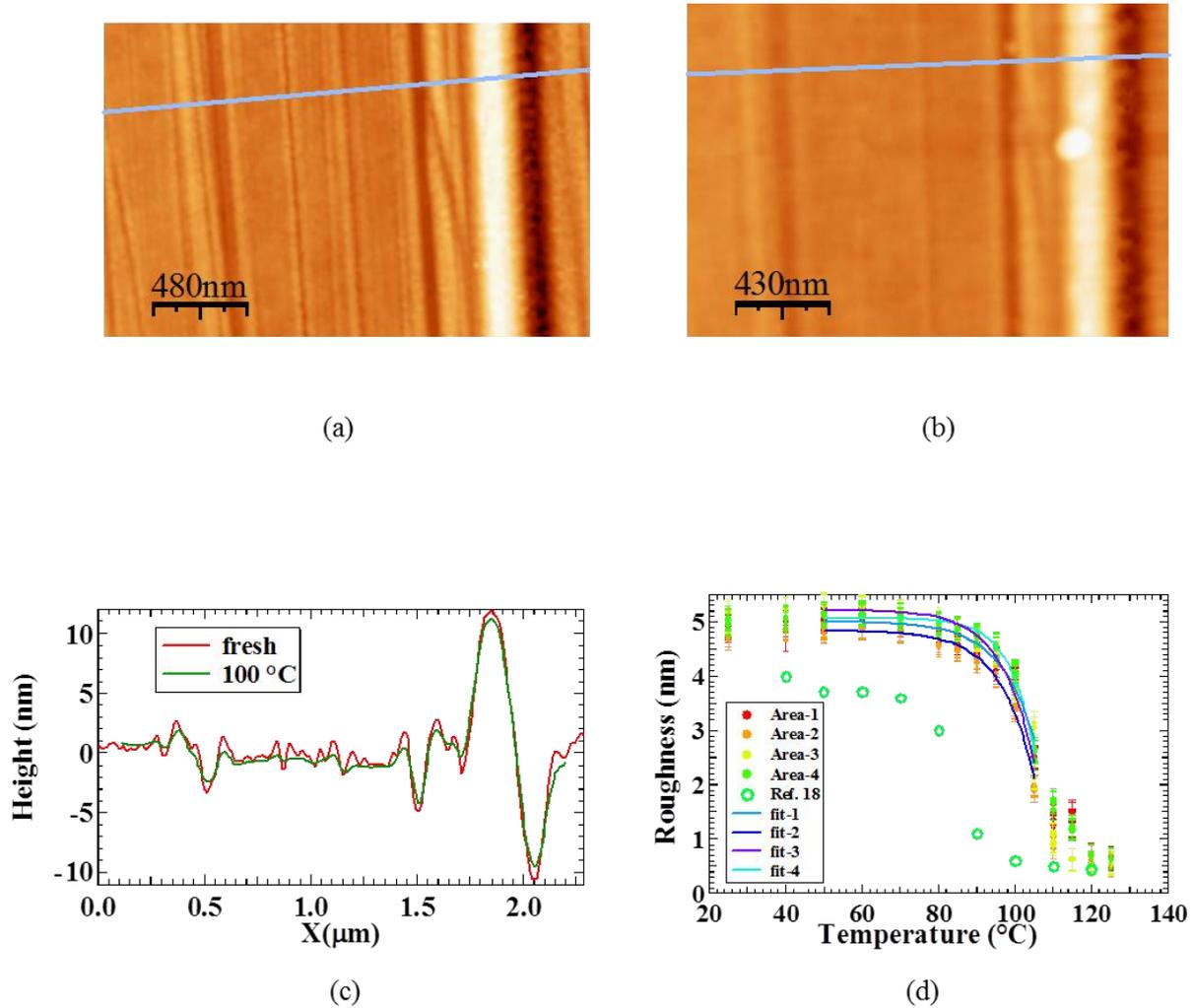

**Figure 1 (a)** SPM image of a fresh rubbed surface of PS; **(b)** SPM of the same area of (a) after the annealing process described in the text; **(c)** The line scan profiles taken from the straight lines in the images in (a) and (b); **(d)** The surface roughness as a function of annealing temperature. The annealing time at each temperature was 30 minutes. The curves are best fits that are described in the text. The data from Ref. 23 are also plotted for comparison.

Figure 2 shows the isothermal temporal evolution of the average roughness at several temperatures. A fixed area of typically 2 x 2 μm$^2$ in size of a fresh sample was analyzed for each temperature. The temperature of the sample was raised directly from 20 °C to the desired temperatures. It is seen that the reduction of roughness became slower at longer time. The same behavior was observed for individual ridges and ditches, and in Ref. 23.

Figure 3 shows the temporal evolution of the roughness of a single sample that underwent consecutive annealing from 80 °C to 100 °C at 5 °C interval for up to 14.5 hours at each temperature. At 80 °C the roughness decreased with time in the same way as in Fig. 2, as expected. At 85 °C, the roughness did not drop up to 1.5 hours of annealing, and only after 4.5 hours did the roughness decrease, unlike the corresponding data in Fig. 2. The same phenomenon can be seen at 90 °C and 95 °C as well. Compared with the data in Fig. 2, there was a 'time delay' in the decrease of roughness at a higher temperature after the sample had been annealed at a lower temperature for a prolonged time. Similar 'delay' effect has been reported in the relaxation of rubbing induced birefringence in PS [21].

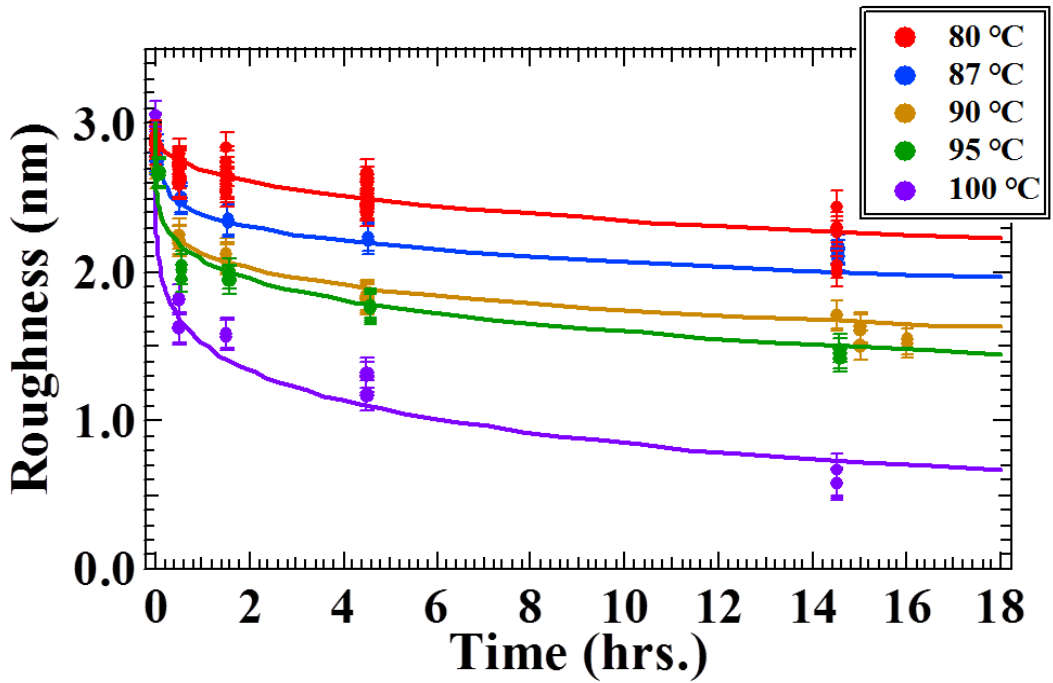

**Figure 2** The isothermal temporal evolution of the surface roughness of fresh surfaces at several temperatures. One fresh sample was used for each set of data at a fixed temperature. The curves are best fits described in the text.

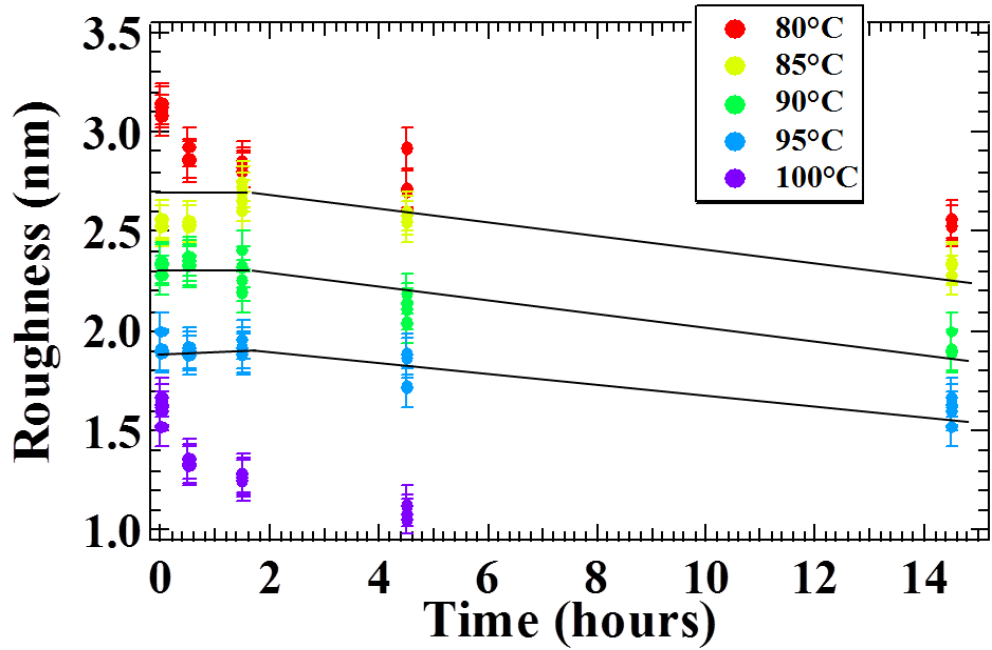

**Figure 3** The isothermal temporal evolution of the surface roughness of a single sample at several temperatures. Notice the 'time delay' phenomenon at 85°C, 90°C, and 95°C, which is absent in Fig. 2. The lines are for guiding the eyes only.

The main driving force to flatten a rough polymer surface is the Laplace pressure $P = \gamma_0/\rho$ [23], where $\gamma_0 = 0.035$ N/m is the surface energy of PS [15, 24], and $\rho$ is the surface curvature in the direction normal to the surface. In the case of an indent of radius $w$ and depth $h$, the effective radius is approximately given by $\rho = \frac{w^2}{4h}$, assuming that the arc passes through the middle point of the bottom and the edge of the indent. The same applies to a long ditch of width $w$ and depth $h$, except that now the relaxation occurs only perpendicular to the ditch. The LP becomes

$$P = 4\gamma_0 h/w^2 \qquad (1).$$

Typical morphology in Fig. 1 ($h = 5$ nm, $w = 100$ nm) gives rise to LP = $7.0 \times 10^4$ N/m$^2$, and varies by a factor of $< 4$ across the surface. This is about 1/1000 of the yield stress of PS [25], so the relaxation is well below the yielding limit [26].

The reduction of roughness on the surface is given by, according to conventional creep experiments,

$$C = PJ(T,t) \qquad (2),$$

where $C$ is the 'relative deformation' with the freshly rubbed surface as the reference, $P$ is the Laplace Pressure, $t$ is the time and $J(T, t)$ is the SEC, which is a sensitive function of temperature $T$. As the width $w$ remained unchanged the average LP is proportional to the roughness, and the 'relative deformation' is proportional to the change of roughness. The temperature variation patterns in the experiments were designed such that out of the three variables, one was kept constant so that the relationship between the other two could be examined. In Fig. 1, time $t$ was fixed, while in Fig. 2 temperature $T$ was fixed. In Fig. 3 after 80 °C the LP was fixed within the 'time delay' period.

Consider the case at 85 °C in Fig. 2 as an example. The reduction of roughness in the first 1800 s is about 0.6 nm. By using the total thickness of 62 nm one obtained a relative change along the film depth of about 1 %. Since the LP is only $7.0 \times 10^4$ Pa, it implies that the average compliance over the entire film depth is about $1.5 \times 10^{-7}$ Pa$^{-1}$, much higher than that of the glassy state ($J_G = 8.5 \times 10^{-10}$ Pa$^{-1}$). If the initial roughness (3.3 nm) is taken as reference the relative change is then 20 %, and the compliance is close to that of the rubbery plateau ($J_R = 6.1 \times 10^{-6}$ Pa$^{-1}$).

In conventional creep experiments, the deformation of the specimen creates internal stress that counter balances the external force until reaching the balance point before the rubbery plateau is reached over a long period. For rubbed PS surface the only 'counter balance' is the reduction of LP due to the reduction of roughness. The roughness drops initially within 1800 s, followed by a drastic slow down. Such behavior cannot be accounted for by a film with uniform compliance that remains constant or *increases* with time, because it cannot explain why under nearly the same LP (reduced by 20 % for the case at 85 °C), the reduction of roughness becomes much slower after the initial 1800 s. It also cannot explain the results in Fig. 3, where under about the same LP, the reduction of roughness did not take place within the first 3 hours when the temperature was raised to 85 °C and above. In other words, after a partial reduction of roughness, the surface became much stiffer. Such phenomenon strongly implies that the compliance depends on the depth from the surface. In Ref. 23, the initial relaxation was attributed to the relaxations of deformed segments, which is consistent with the non-uniform compliance.

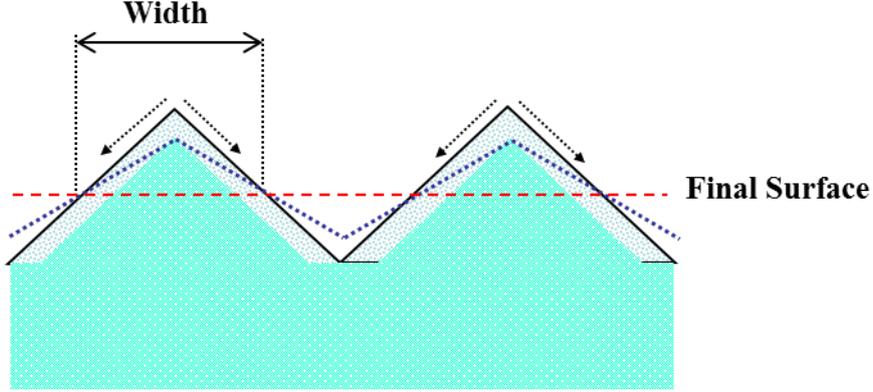

**Figure 4 Schematic illustration of the surface morphology evolution. The fresh surface is marked by the solid lines, with the lightly shaded region underneath being the mobile layer. The surface after partial relaxation is marked by the blue dashed lines. The fully relaxed surface is marked by the red dashed line.**

The observed relaxation behavior can be summarized schematically in Fig. 4. There is a surface layer with a depth dependent $T_g$. At elevated temperatures (~ 85 °C) the mobility of a layer (lightly shaded region) on the original fresh surface becomes high enough that it moves downhill under the LP, as indicated by the arrows. When the movement is completed, the exposed deeper layer on the hilltop maintain its $T_g$, even though it is now the very top layer, and further relaxation becomes much slower (by ~ 50 times). Such behavior exemplifies the seemingly contradictory features of polymer surfaces. The original top surface layer is much more mobile than the bulk, and the mobility decreases drastically in the region underneath the surface. On the other hand, if the newly exposed surface layer was as mobile as the original one, the reduction of roughness would continue with the same speed as in the initial 1800 s until the LP is much reduced. This is in clear contradiction with the experiments.

The reason why the partially relaxed surface is not as mobile as the original one may lie in the following argument. Surface mobility is enhanced because of the enrichment of chain ends [8 – 10] and free volume [3, 4]. However, because the remaining portion of a surface molecule is still anchored underneath [1], the surface segments can only be displaced by a finite distance limited by the length of the segments. Based upon such physical picture, we propose a depth-dependent SEC that takes the form [15]

$$J(T,\xi,t) = J_G + (1 - e^{-(t/\tau(T,\xi))^\beta})(J_R - J_G) \qquad (3),$$

where

$$\tau(T,\xi) = \tau_0 e^{E(\frac{1}{T+273} - \frac{1}{\xi+273})} = \tau_0 e^{\varepsilon(\xi - T)} \qquad (3A)$$

is the characteristic time in Arrhenius form, $\varepsilon \equiv \dfrac{E}{R(273+T)(273+\xi)}$, $E$ is the activation energy, $T$ is the temperature in the same unit of °C as $\xi$, and $\xi$ is essentially the depth ($y$) dependent T$_g$. It should be emphasized that $\xi$ depends only on the *original* depth $y$ of the region. The $\xi$ of a region will remain the same even when the depth of the region is reduced after the polymer segments above it have moved away during the relaxation process. This specific assumption is made in accordance with the experimental observation in Fig. 2 & 3. It is justified because the original mobile segments are still in the vicinity while the remaining portions of the molecules remain anchored underneath.

We now present the quantitative analysis of the experimental results, assuming that the relaxation at a particular depth is complete within 1800 s when the temperature reaches the T$_g$ of the depth. The dependence of $\xi$ on $y$ is obtained from the experimental data in Fig. 1. The reduction in roughness is found to increase exponentially with temperature, up to the point where the reduction of the LP can still be ignored (~ 2 fold) as compared to the change of mobility with depth (~ 100 fold). According to Eq. (3), at a given $T$ only the layer down to the depth of $\xi(y) = T$ has relaxed, so the depth is equal to the amount of reduction of roughness. The curves in Fig. 1(c) are the fittings to a single exponential function

$$y = h_0 e^{\gamma(\xi - T_0)} \qquad (4)$$

The results are $\gamma = 0.11 \pm 0.01$ °C$^{-1}$, $h_0 = 2.1 \pm 0.3$ nm, and $T_0 = 102 \pm 2$ °C. Depth $y$ becomes tens of nanometers when $\xi$ reaches the bulk T$_g$, as expected. Replacing $\xi$ by $T$ in the equation, one obtains the thickness of the surface mobile layer at a given temperature. At 80 °C the mobile layer thickness is about 0.2 nm, which may be due to the wiggling motion of segments at the surface.

The parameter $\varepsilon$ can be estimated using the data in Fig. 3. The fact that after annealing at 80 °C for 14.5 hours, the roughness at 85 °C did not drop up to 1.5 hours of annealing, and only after 4.5 hours did the roughness decrease indicates, according to Eq. (2), that $\frac{14.5}{4.5} \leq \frac{\tau(80°C, \xi)}{\tau(85°C, \xi)} \leq \frac{14.5}{1.5}$. Similarly, from the experimental 'time delay' at higher temperatures we obtain $\frac{14.5}{4.5} \leq \frac{\tau(85°C, \xi)}{\tau(90°C, \xi)} \leq \frac{14.5}{1.5}$, $\frac{14.5}{4.5} \leq \frac{\tau(90°C, \xi)}{\tau(95°C, \xi)} \leq \frac{14.5}{1.5}$, and $\frac{14.5}{0.5} \leq \frac{\tau(95°C, \xi)}{\tau(100°C, \xi)}$. This leads to $0.23 \leq \varepsilon \leq 0.45$ in the temperature range (80 °C, 95 °C) and $0.67 \leq \varepsilon$ at the temperatures > 95 °C. It is seen that the value of $\varepsilon$ increases slowly with temperature, and the corresponding active energy $E$ is in the range of the α-relaxation.

We now examine the isothermal temporal evolution of roughness starting from fresh surfaces with roughness $H_0$. Assuming that the relaxation can be regarded as complete when the compliance reaches $J_R$, the relaxation of roughness $H$ can be numerically modeled as the 'melt down' of a rod of length $H(t)$ with a position dependent T$_g$ along the length. The small section $dH$ at depth $y$ (as measured from the top of the rod) decays according to $dH = dy e^{-(t/\tau(T,\xi(y)))^\beta}$, i. e., the temporal evolution of the roughness is modeled as the reduction of each small length $dH$ following the extended exponential function of $t$ with a depth $y$ dependent time constant $\tau(T, \xi(y))$. Then the roughness $H(t)$ is given by

$$H(t) = \int_0^{H_0} e^{-(t/\tau(T,\xi(y)))^\beta} dy \qquad (5).$$

Using Eq. (4) we have

$$H(t) = H_0 \int_0^{H_0/h_0} e^{-(t'z^{\varepsilon/\gamma})^\beta} dz \qquad (6),$$

where $t' \equiv \dfrac{t}{\tau_0} e^{\varepsilon(T-T_0)}$ is the reduced time. The dominant feature of Eq. (6) is the 'master curve' characteristics which is insensitive to $H_0$ and other parameters, and the main time dependence is contained in the shifted factor $e^{\varepsilon T}$. By properly shifting the experimental $H \sim t$ data horizontally on the logarithmic time axis to the 'master curve' one can obtain the shift factor $e^{\varepsilon T}$. The curves in Fig. 2 are the best fits to the experimental data. The $\varepsilon$ values obtained from the fitting are plotted in Fig. 5. They are consistent with the experimental 'time delay' effects in Fig. 3. The corresponding activation energy $E$ is between 250 to 500 KJ/mol, i. e., the relaxation is of the nature of α-relaxation. This is expected because the relaxation requires significant displacements of segments. The activation energy increases with temperature. Since the surface mobile layer thickness also increases with temperature, this may indicate that the activation energy is also depth dependent, being about half the bulk α-relaxation value at the surface. This is consistent with the findings by others [8 – 10]. The final characteristic time for SEC in Eq. (3) is

$$\tau(T,y) = \tau_0 \left(\frac{y}{h_0}\right)^{\frac{\varepsilon}{\gamma}} e^{\varepsilon(T_0-T)} \qquad (7).$$

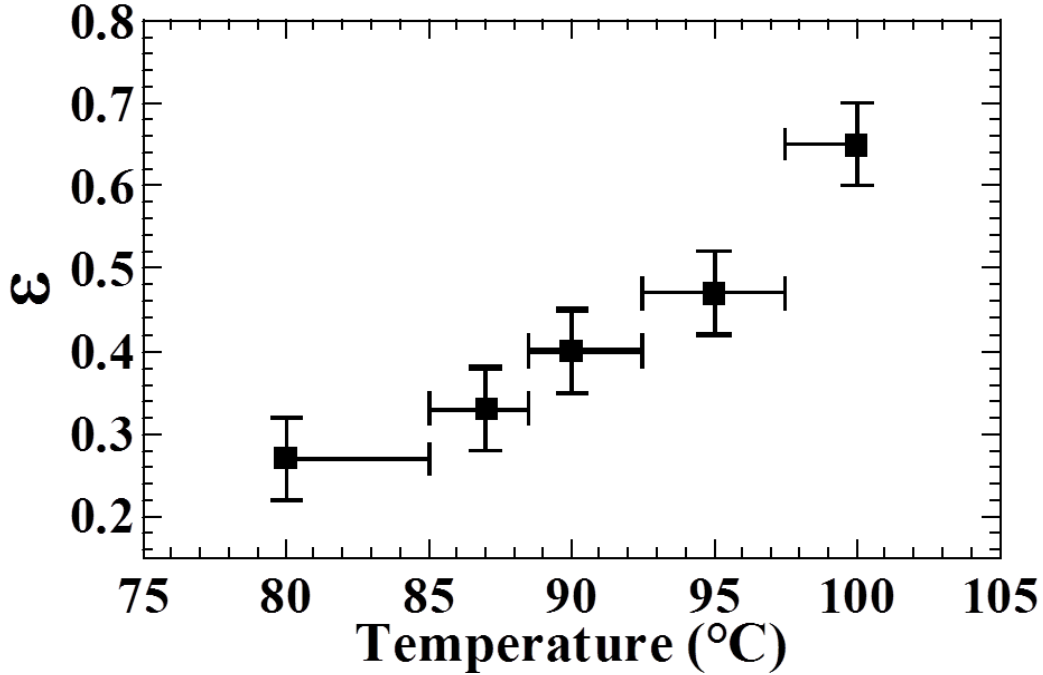

**Figure 5 The parameter $\varepsilon$ as a function of temperature obtained from the best fit to the experimental data in Fig. 2.**

We now examine the results reported in Ref. 23. There the initial aspect ratio is about 1 and the radius is ~10 nm. In our case, the aspect ratio is about 0.05 while the width is about 100 nm. The LP is about 200 times smaller in our case. Using the $\varepsilon$ value at 80 °C the compliance that is 200 times higher should occur at about 20 °C lower in temperature. As seen in Fig. 1, the decrease in roughness in Ref. 23 indeed occurred at about 20 °C below the corresponding temperature in our case.

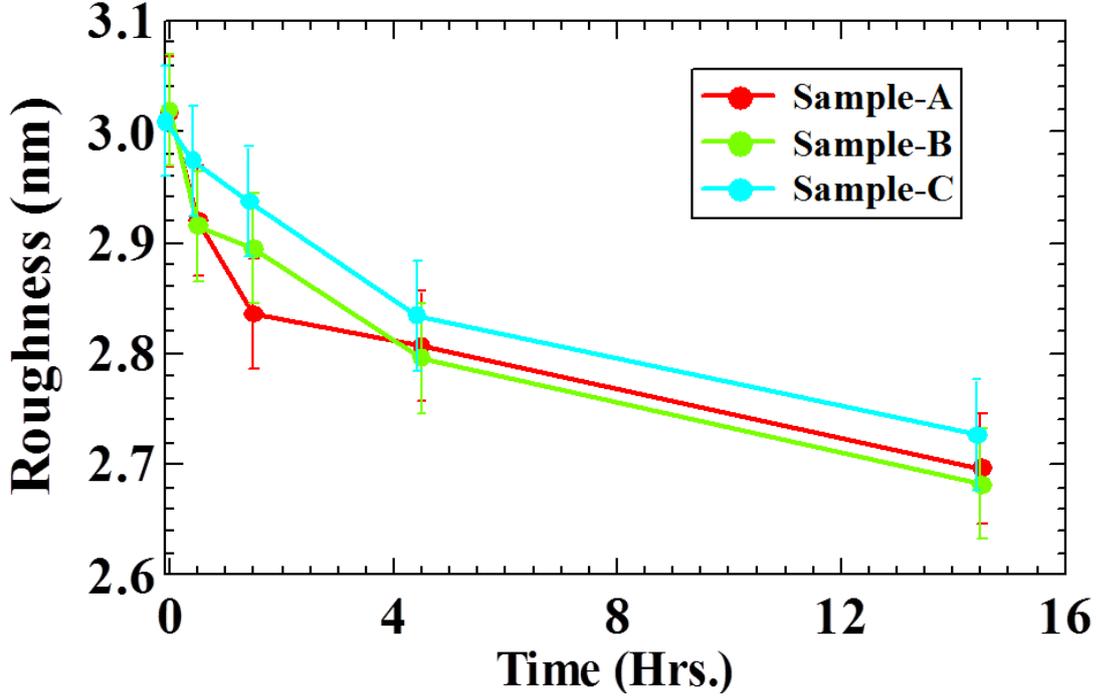

**Figure 6 The isothermal temporal evolution at 85°C of the surface roughness of three samples aged according to the description in the text.**

Although the results reported here are obtained from highly deformed surfaces, we nevertheless argue that they reflect the properties of PS surface in general. This is based on the following reasons. (i) The rubbing induced strain field is expected to be more random, and highly unlikely to induce uniform relaxation behavior shown in Fig. 1. Furthermore, rough surface prepared in a different way [23] showed consistent relaxation behavior with the rubbed surface, although the strain fields in the two cases are expected to be very different. (ii) The surface is already near the rubbery plateau at 80 ºC. But for bulk samples immediately after quench the rubbery plateau will not be reached at 80 ºC within 1800 s. Such rubbery layer therefore cannot be due to the mechanical 'rejuvenation' [25]. (iii) It is well known that deformed bulk samples exhibit much higher segmental mobility as in the cases by quenching from above $T_g$ [25]. However, surface layers (~ 10 nm thick) age quickly, due to free volume diffusion [27, 28]. We have also conducted aging study on surface relaxations. The most relevant results are presented here, and the more extensive ones are reported in [29]. Figure 6 shows the isothermal temporal evolution of roughness at 85 ºC of three samples. The first sample (Sample-A) was measured within one hour after rubbing. The second sample (Sample-B) was aged for 2 days at 50 ºC after rubbing. The third sample (Sample-C) was aged at 20 ºC for 7 days after rubbing before it was measured. As can be seen, the temporal evolutions of the three samples are identical within experimental uncertainties. In particular, since the aging times of Sample-A and Sample-C differ by about 170 times, even at a shift rate of 0.6 [25], the bulk creep compliance curve of Sample-C should have shifted by a factor of $170^{0.6} = 21.8$ relative to Sample-A. Had the SEC followed the bulk creep compliance, it would have taken 21.8 hours for the roughness of Sample-C to decrease by the same amount as Sample-A at 1.0 hour. No detectable drop in its roughness should have been observed within the experimental time span of 14.5 hours. The fact that the surfaces of all three samples behaved the same is consistent with a surface layer in which aging is complete in an hour at 20 ºC, and the deformed surface layer recovers its original viscoelastic properties, which are revealed in this work.

We conclude that on PS surface there is a heterogeneous surface layer with enhanced mobility at the surface and drastic decreasing mobility with depth, while having temperature dependence close to the α-relaxation. The segments in the layer can only be displaced by a finite distance comparable to the depth. Such a surface layer indicates that near the surface, motions of different length scales exhibit different $T_g$ suppressions. The smaller the scale is, the more suppression. Ellison and Torkelson reported reduction of $T_g$ by nearly 30 °C in a surface layer of 14 nm thick [9], which is much larger than ours. This is because the random motion of segments that affects the fluorescence intensity involves a much smaller length scale than the nanometer-size movement reported here. Other seemingly contradictory experimental results can also be qualitatively explained. The properties that are related only to the small local segmental motions will exhibit much enhanced mobility. These include the free volume probed by positrons and muons [3, 4], the surface friction [8 – 10], the thermodynamic properties [11 – 15], and physical aging. Properties requiring large displacements of the segments, such as the sinking of metallic spheres [3, 15], or involving the entire film much thicker than the surface layer [16], will exhibit less mobility enhancement or suppression of $T_g$. While the development of a unified microscopic theory for the dynamic properties of polymer surfaces is still under way [1], the results reported here provide a concrete basis for quantitative tests of the emerging theoretical models.

**Method**

Mono-dispersed PS with molecular weights 102 Kg/mol ($M_w/M_n$ < 1.1) in toluene solution were spin coated on silicon wafers with 20 nm of thermal $SiO_2$ to form 62 nm films. The samples were then annealed in vacuum at 150°C for several days before slowly cooled down (< 0.1 ºC/min) to room temperature. Rubbing was done at normal pressure of 9 g/cm$^2$ and at a constant speed of 1 cm/s. The rubbed samples were then kept at room temperature for at least 24 hours before their surface topography were measured on a scanning probe microscope (SPM) in tapping mode. An on-stage heater provided the controlled annealing. The sample was then cooled down to room temperature, and its surface topography was measured. The heating and cooling to and from the set temperature usually took about 5 minutes. The micrographs of nearly the same 5 × 5 μm$^2$ area were recorded for each set of measurements. Using recognizable topological features an identical area of typically 2 × 2 μm$^2$ was selected for analysis.

**References**

[1] P.G. de Gennes. *Eur. Phys. J. E* **2**, 201–205 (2000)
[2] Mataz Alcoutlabi and Gregory B McKenna. *J. Phys.: Condens. Matter* **17** (2005) R461–R524
[3] G. B. DeMaggio, W. E. Frieze, and D. W. Gidley, Ming Zhu, H. A. Hristov, and A. F. Yee. *Phys. Rev. Lett.* **78**, 1524 - 1527 (1996)
[4] F. L. Pratt, T. Lancaster, M. L. Brooks, and S. J. Blundell, T. Prokscha, E. Morenzoni, A. Suter, H. Luetkens, R. Khasanov, R. Scheuermann, and U. Zimmermann, K. Shinotsuka and H. E. Assender *Phys. Rev.* **B72**, 121401(R) - 121404 (2005)
[5] Sharp J S and Forrest J A. *Phys. Rev. Lett.* **91** 235701 (2003)
[6] J.A. Forrest, K. Dalnoki-Veress, J. Stevens, J.R. Dutcher. *Phys. Rev. E* **56**, 5705 - 5716 (1997)
[7] Shin Kawana, Richard A. L. Jones. *Phys. Rev.* **E63**, 21501 - 21506 (2001)
[8] Mikhail Yu. Efremov, Eric A. Olson, Ming Zhang, Zishu Zhang, and Leslie H. Allen. *Phys. Rev. Lett.* 91, 85703 (2003)
[9] Christopher J. Ellison and John M. Torkelson. *Nature Mat.*, **2**, 695 – 700 (2003)
[10] Kajiyama T, Tanaka K, Satomi N, Takahara A. *Science & Technology of Advanced Materials,* **vol.1**, 31 - 35 (2000)
[11] Xiqun Jiang, Chang Zheng Yang, Tanaka K, Takahara A, Kajiyama T. *Physics Letters A,* **281***, 363 – 367 (2001)
[12] Keiji Tanaka, Daisuke Kawaguchi, Yasuyuki Yokoe, Tisato Kajiyama, Atsushi Takahara, Seiji Tasaki. *Polymer* **44** (2003) 4171–4177
[13] Jörn Erichsen, Jörn Kanzow, Ulrich Schurmann, Kai Dolgner, Katja Günther-Schade, Thomas Strunskus, Vladimir Zaporojtchenko, and Franz Faupel. *Macromolecules 37,* 1831-1838 (2004)
[14] J. H. Teichroeb and J. A. Forrest. *Phys. Rev. Lett.* 91, 16104 (2003)
[15] S. A. Hutcheson and G. B. McKenna. *Phys. Rev. Lett.* **94,** 076103 (2005)
[16] P. A. O'Connell and G. B. McKenna. Polymer Physics Workshop 2006, Suzhou, China
[17] X. P. Wang, O. K. C. Tsui, and X. D. Xiao. *Langmuir* **18**, 7066 - 7072 (2002)
[18] Liu, Y.; Russel, T. P.; Samant, M. G.; Stohr, J.; Brown, H. R.; Cossy-Favre, A.; Diaz, J.. *Macromolecules*, **30**, 7768-7771 (1997)
[19] Zongyi Qin, Yonghai Chen, K. P. Shiu, and Z. Yang. *Macromolecules*, **37***, 3378-3380 (2004)
[20] Alexander D. Schwab and Ali Dhinojwala. *Phys. Rev. E.*, **67**, 21802 (2003)
[21] K.P. Shiu, Zongyi Qin, and Z. Yang. *Eur. Phys. J.* E 17, 139 - 147 (2005)
[22] Günter Reiter, Moustafa Hamieh, Pascal Damman, Séverine Sclavons, Sylvain Gabriele, Thomas Vilmin and Elie Raphaël. *Nature Materials* **4**, 754 (2005)
[23] Tobias Kerle, Zhiqun Lin, Ho-Cheol Kim, and Thomas P. Russell. *Macromolecules*, **34***, 3484-3492 (2001)
[24] *Polymer Interface and Adhesion*, Souhang Wu, p 68 – 70, (1982, Marcel Dekker, Inc., New York, Basel)
[25] H.G.H. van Melick, L.E. Govaert, B. Raas, W.J. Nauta, H.E.H. Meijer. *Polymer* **44**, 1171 (2003)
[26] G. Reiter. *Eur. Phys. J. E* **8**, 251–255 (2002)
[27] S. Kawana, and R.A.L. Jones. *Eur. Phys. J.* **E 10**, 223–230 (2003)
[28] Kokou D. Dorkenoo, Peter H. Pfromm. *Journal of Polymer Science: Part B: Polymer Physics,* **Vol. 37**, 2239–2251 (1999)
[29] K.P. Shiu, Zongyi Qin, and Z. Yang, *Eur. Phys. J. E* 24, 385 – 397 (2007)